\documentclass{aa}

\usepackage{graphicx}
\begin{document}

\title{A New Approximation Of ECM Frequencies}  

\author{A. Stupp\inst{1}}

\institute{School of Physics and Astronomy,Tel-Aviv University, Israel}

\maketitle

\begin{abstract}
  We investigate wave amplification through the Electron 
Cyclotron Maser mechanism. 
  We derive a semi-analytic approximation formula giving the frequencies at 
which the absorption coefficient is negative. The coefficients still need to be
computed to obtain  the largest, and therefore the dominant, coefficient.
\end{abstract}


\begin{acknowledgements}
 I would like to thank Dr. R. Ramaty for his time and comments.
\end{acknowledgements}

\section{Introduction}
  The observation of millisecond microwave spikes from the Sun 
has been interpreted as gyrosynchrotron maser emission since the
late 1970`s (\cite{Wu}; \cite{Holman}). The spikes exhibit a
duration of a few milliseconds to some tens of milliseconds,
and a narrow bandwidth ($\approx 1\%$ or a few MHz). The characteristic high 
brightness temperature ($10^{15-18}\ K$) deduced for the spikes strongly
suggests maser action (\cite{BenzBook}). \par
  Several papers have explored the possibilities of cyclotron maser. 
Melrose and Dulk (\cite{D+M}) have used the semi-relativistic 
approximation to derive approximate formulae for the frequency of the largest 
growth rate, and for the growth rate. Winglee (\cite{Winglee}) 
approximated the effects of the ambient plasma temperature on the maser emission. 
Aschwanden (\cite{Aschwanden a}, \cite{Aschwanden b}) 
followed the diffusion of electrons into the loss-cone as a result of the maser 
emission, and therefore the self-consistent closure of the loss-cone as the energy is
transferred to the maser. 
Kuncic and Robinson (\cite{Kuncic})  performed ray-tracing in a model loop with
dipole field, and concluded that maser emission can escape from lower
levels.  \par
   Some papers have also attempted to derive analytical approximate formulae
for the frequencies of the maser.
   Dulk and Melrose (\cite{D+M}) concluded that the maser emission is at, 
or near, $90$ degrees to the magnetic field and the maximum growth rate they 
find is very near the cyclotron frequency, or at the second harmonic. 
   Hewitt and Melrose (\cite{Hewitt}) derive the conditions for solutions
and the number of solutions, but not the frequencies themselves. \par
     All previous studies concentrated on the {\it growth rate}
of the amplified wave, we use the {\it absorption coefficient} approach
of Ramaty (\cite{R}) and calculate the frequencies at which the
absorption coefficient is negative. \par

\section{Gyrosynchrotron Absorption}
 In studies of microwave emission from solar flares it is customary to assume that
the plasma is composed a non-thermal population of energetic elecrons, and a thermal
population of much higher density, but much lower energy. The thermal
plasma is assumed to determines the propagation of waves. This plasma can
be described by the cold plasma approximation using magnetoionic theory (\cite{MelroseB}). \par
  ECM studies usually compute the growth rate, while we 
follow Ramaty (\cite{R}) in using the absorption coefficient, but
for the purposes of this article the detailed equations are unimportant.
Both the growth rate equation and the absorption coefficient equation include 
a $\delta$-function from which the important resonance condition is derived
\begin{equation}
\label{resonance}
{{s\omega_B}\over \gamma}=\omega (1-n_\pm \beta\cos(\theta)\cos(\phi))
\end{equation}
where $\omega_B=eB/m_e/c$ is the cyclotron frequency, $\omega$ is the
emission frequency, $s$ is an integer,
$\theta$ is the angle of the wave vector ${\bf k}$ to the magnetic field,
$\phi$ is the angle between the velocity of the electron and the magnetic field (the
pitch angle), and $\gamma$ and $\beta$ are the usual Lorentz factor and velocity. 
The important factor in equation \ref{resonance} is $n_\pm$ the refraction index 
for the Ordinary Mode (OM or '+'), and the eXtra-Ordinary mode (XO or '-').
The refraction index is a function of the ambient density, of the cyclotron frequency,
of the emission frequency, and of the angle to the magnetic field
\begin{eqnarray}
n_\pm^2=&& 1 + \\
&&\nonumber
 { {2\nu_p^2(\nu_p^2-\nu^2)}\over 
{-2\nu^2(\nu_p^2-\nu^2)-\nu^2\nu_B^2\sin^2(\theta) \pm 
\sqrt{\Delta} }}\\
&&\nonumber
\Delta=\nu^4\nu_B^4\sin^4(\theta)+
4\nu^2\nu_B^2(\nu_p^2-\nu^2)^2\cos^2(\theta)
\end{eqnarray}
Where $\nu_p$ is the plasma frequency and the rest are as in equation
\ref{resonance}. \par
  If $\theta \neq \pi/2$ the $\delta$ function is transformed to 
\begin{eqnarray}
&&\delta\left(\nu-{{s \nu_B}\over{\gamma}}-n_\pm \nu \beta \cos(\phi) \cos(\theta) \right) =\\
&&\nonumber
\delta\left(\cos(\phi)-{{1-{{s\nu_B}\over{\gamma\nu}}}\over{n_\pm\beta\cos(\theta)}}\right)
{{1}\over{n_\pm\nu\beta\cos(\theta)}}
\end{eqnarray}
This transformation gives us the pitch angle $\phi_s$ of an electron emitting in direction
$\theta$ a wave of frequency $\nu$, where  
\begin{equation}
\label{cosphi_s}
\cos(\phi_s)={ {1- { {s\nu_B}\over{\gamma \nu}}}\over{n_\pm \cos(\theta) \beta} }
\end{equation}
\par
  The absorption coefficient after the integration over the
pitch angle is (\cite{R})

\begin{eqnarray}
\label{gyroe}
&&k_\pm \left( {\nu ,\theta } \right) = 
{1 \over B}4\pi ^2e
{{2\pi } \over {\left| \cos \left( \theta  \right) \right|}}
{{\nu _B} \over \nu }{1 \over {n_\pm ^2}} \times \\
&&\nonumber
\int_1^\infty  d\gamma {{u\left( \gamma  \right)} \over \beta }
\sum\limits_{s=s_1}^{s_2}
{{g\left( {\phi _s} \right)} \over {1+T_\pm ^2}}
\Bigl[ \beta \sin \left( {\phi _s} \right)J_s^\prime\left( {x_s} \right)+ \\
&&\nonumber
{{L_\pm}\over{n_\pm}}J_s\left(x_s \right)+ 
T_\pm \left({{\cot \left( \theta  \right)} \over {n_\pm }}-
{{\beta \cos \left( {\phi _s} \right)} \over {\sin \left( \theta \right)}}
 \right)J_s\left( {x_s} \right) \Bigr]^2\times\\
&&\nonumber
\Bigl[ {{-\beta \gamma ^2} \over {u\left( \gamma  \right)}}      
{d \over {d\gamma }}
{{u\left( \gamma  \right)} \over {\beta \gamma ^2}}+
{{n_\pm \beta \cos \left( \theta  \right)-\cos \left( {\phi _s} \right)} \over
{\gamma \beta ^2\sin \left( {\phi_s} \right)}} \times \\
&&\nonumber
{1 \over {g\left( {\phi _s} \right)}}
{{dg\left( \phi  \right)} \over {d\phi }} \Bigr]
\end{eqnarray}
Where $J_s$ are Bessel functions, with the argument 
$x_s=(\nu/\nu_B) n_\pm \gamma \beta \sin(\theta) \sin(\phi)$. 
  The cold plasma parameter $T_\pm$ is the ratio of the polarization ellipse axis, which 
describes the transverse part of the polarization vector, and the parameter $L_\pm$ describes
the longitudinal part of the polarization vector. Equations for $T_\pm$ and $L_\pm$
can be found in Melrose (\cite{MelroseB}).
The range of the summation over the Bessel functions is
\begin{equation}
\label{s1-s2}
s_{1,2}={{\nu \gamma}\over{\nu_B}} \left( 1 \pm n_\pm \beta \cos(\theta) \right)
\end{equation}
And $u(\gamma)g(\phi)$ is the distribution function of the emitting electrons. \par
  We assume the distribution function in equation \ref{gyroe} can
be put in the form
\begin{displaymath}
f({\bf p})d^3p=2\pi u(\gamma)g(\phi)d\gamma d(\cos(\phi))
\end{displaymath}
And take for the distribution function of the non-thermal electrons a power-law
\begin{equation}
\label{powerlaw}
  u(\gamma)=A\cdot (\gamma-1)^{-\delta}
\end{equation}
which is normalized to $n_{hot}$ the number density of the energetic electrons.
  For the pitch angle dependence in the energetic distribution we use an
ideal loss-cone which describes an isotropic distribution for pitch
angles above some $\alpha$, no electrons below another angle
$\alpha - \Delta \alpha$, and a linear in $cos(\phi)$ decrease within
the $\Delta \alpha$ interval (\cite{D+M}), or :
\begin{equation}
\label{losscone}
g(\phi)=\cases{ A, &$\cos(\phi) < \cos(\alpha)$ \cr
               A\times{{cos(\alpha-\Delta\alpha)-cos(\phi)} \over {cos(\alpha-\Delta\alpha) - \cos(\alpha)}}, 
&$\cos(\alpha) < \cos(\phi) < $\cr
&$cos(\alpha-\Delta\alpha)$ \cr
                        0, &$cos(\alpha-\Delta\alpha) < \cos(\phi)$ \cr}
\end{equation}
Where $A$ is a constant which
depends on the loss cone parameters, and for an idealized loss
cone as above is $A =1 / \pi /(2+\cos(\alpha)+cos(\alpha-\Delta\alpha))$.
With this constant the integral is normalized so that
$2\pi \int_{-1}^1 g(\phi) d(\cos\phi)= 1$. \par

\section{Conditions for Maser}
The Electron-Cyclotron-Maser occurs when the absorption coefficient
is negative. The resulting amplification of the radiation is referred
to as {\it maser}. \par
   Inspection of equation \ref{gyroe} shows that the absorption coefficient
for a distribution which is isotropic in pitch angle can only be negative if
the slope of $u(\gamma)/\gamma^2/\beta$ is positive. This is the equivalent
of an increase of the fast particles with energy, a distribution which is
not likely to occur. The other possibility for amplification is when the distribution
is not isotropic in pitch angle. In that case if the derivative of $g(\phi)$ is large 
enough, and there are solutions such that the pitch angle dependent term 
in equation \ref{gyroe} is negative, then the absorption will be negative. 
One such anisotropic distribution is the loss-cone distribution, which we
use. The loss-cone distribution produces negative absorption coefficients
and maser emission.\par

\section {Semi-Analytical Formulae}
  The resonance condition \ref{resonance}
defines an ellipse on the electron velocity plane $(v_{||},v_\perp)$
for a given $s$,$\theta$, and frequency of emission
(\cite{MelroseB}).
The integral over the electron distribution function,
which is necessary to calculate the absorption coefficient,  
is therefore an integral along the ellipses of the 
various $s$ harmonics.  
  In the weakly relativistic case $1/\gamma=1-\beta^2/2$
this ellipse becomes a circle (\cite{D+M}) - the
resonance circle - whose center is on the $v_{||}$-axis (the x-axis) 
at $v_{||}=v_c$, and whose radius is $v_R$ with
\begin{equation}
{{v_c}\over{c}}={{n_\pm \cos(\theta) \nu}\over {s \nu_B}}
\end{equation}
\begin{equation}
{{v_R}\over{c}} = \sqrt{ \left({{v_c}\over c}\right)^2 - {{2(\nu-s \nu_B)}\over{s \nu_B}} }
\end{equation}

  The loss-cone described by equation \ref{losscone} divides
the $(v_{||},v_\perp)$ plane into three regions. Below the line
of angle $\alpha-\Delta\alpha$ to the $v_{||}$-axis there are 
no electrons, above the line of angle $\alpha$ to the $v_{||}$-axis
the distribution is isotropic in angle, and between the two lines the
number density of electrons decreases with angle. \par
\begin{figure}
\resizebox{\hsize}{!}{\includegraphics{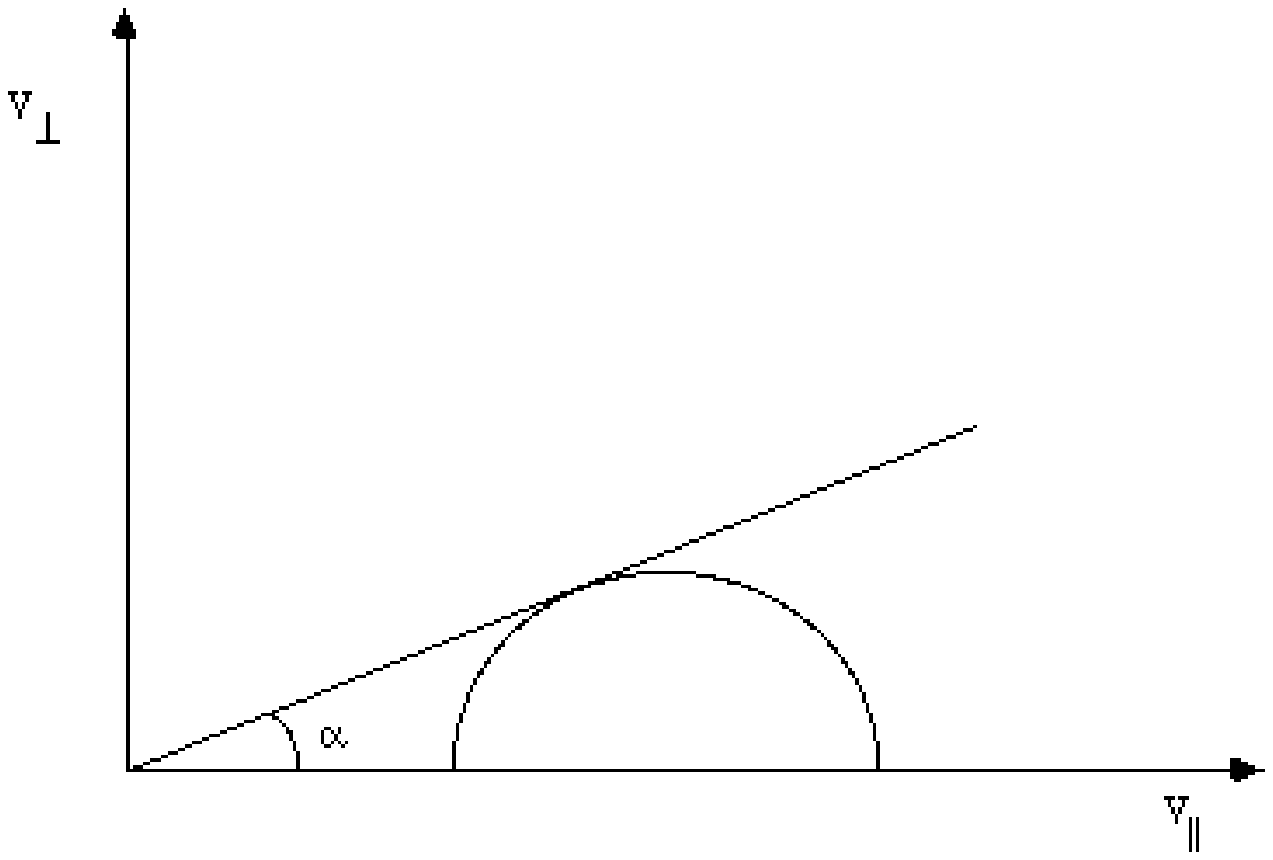}} 
\caption[Loss-cone and resonance circle]
{A loss-cone on the plane $(v_{||},v_\perp)$ is a straight line with 
angle $\alpha$ to the x-axis. The resonance condition defines, with the 
small $\gamma$
approximation, a resonance circle on the plane. The special circle which 
is tangent
to the line of the loss-cone is the circle for the frequency where the 
largest in
absolute value negative absorption occurs. }
\label{losscone-figure}
\end{figure}

  The frequency and angle for which the largest (in absolute value) 
negative absorption is expected
are those for which the resonance circle is tangent to the outer edge
of the loss-cone at $\alpha$ (\cite{D+M}), 
as shown in figure \ref{losscone-figure}.
Integration along this special circle passes only through the two
lower regions
where there are either no electrons, or the distribution changes with angle.
Where there are no electrons the absorption is zero, and where the
distribution changes with angle is where the integrand is most likely
to be negative, and therefore
integration around the special circle is the most likely to
give negative absorption. The special circle also has the
longest possible path through this region, and is
expected to produce the largest (in absolute value) negative absorption
coefficient. \par
  The definition of the special circle results
in an equation for a frequency which is the frequency
where the largest absorption is expected.
This frequency is (\cite{D+M}) 
\begin{equation}
  \label{DM}
{\nu \over {\nu_B}}=s\left(1+{{ n_\pm^2\cos^2(\theta)\cos^2(\alpha)}\over 2}\right)
\end{equation}
  Since even a short path through the positively contributing region of
the $(v_{||},v_\perp)$ plane increases the positive absorption considerably,
the expectation is that there are negative absorption coefficients only for
this special frequency, or at frequencies very close to it. \par
  For the fully relativistic case it is more convenient to consider
the plane $(\gamma,\cos(\phi))$. On this plane the loss cone regions
are the region below the line parallel to the $\gamma$-axis (the $x$-axis) at
$\cos(\phi)=\cos(\alpha-\Delta\alpha)$, where the number density
is independent of the pitch angle, the region above the line
$\cos(\phi)=\cos(\alpha)$, where there are no electrons,
and the region between the lines, where the number density
decreases. On the $(\gamma,\cos(\phi))$ plane the resonance condition
defines a curve which is given by equation \ref{cosphi_s}. \par
\begin{figure}
\resizebox{\hsize}{!}{\includegraphics{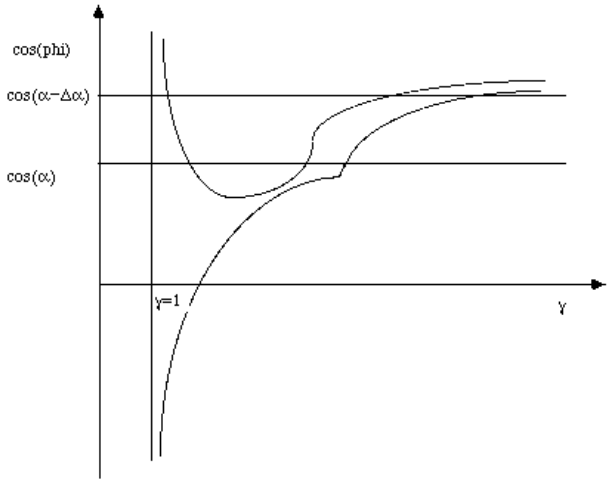}}
\caption[Loss-cone and relativistic resonance curves]
{On the plane $(\gamma,\cos(\phi))$ the solution of the resonance 
condition for the pitch angle (equation \ref{cosphi_s})
can appear
in one of two forms. The first curve goes to $+\infty$ when $\gamma$ 
approaches $1$ from above,
and the second curve goes to $-\infty$ when $\gamma$ approaches $1$. 
For some frequency
the first form is tangent to the $\cos(\phi)=\cos(\alpha)$ outer edge of the 
loss-cone,
and this curve has the largest in absolute value negative absorption 
coefficient. }
\label{cosphi-figure}
\end{figure}

  There are two possible forms to the curve given by equation \ref{cosphi_s},
as shown in figure \ref{cosphi-figure}.
The first form is for $s\nu_B > \nu$. For this case, as $\gamma$ approaches $1$
from above, and assuming $\cos(\theta)>0$, the $\cos(\phi_s)$ solution goes
to minus infinity. The second form of the curve is for $s\nu_B < \nu$, and 
for this form the $\cos(\phi_s)$ solution goes to positive infinity as the
kinetic energy approaches zero.
Both curves go asymptotically to $[n_\pm \cos(\theta)]^{-1}$ as $\gamma$
goes to infinity.
The second form  curve (for $\nu > s \nu_B$) undergoes a change in the sign of the
slope at some energy. For that curve the value of $\cos(\phi_s)$ ceases to decrease as
$\gamma$ increases (negative slope), and starts to increase (positive slope), as
figure \ref{cosphi-figure} illustrates.  \par
  For the fully relativistic case the equivalent curve to
the special resonance circle is a curve with the second form, which on
the $(\gamma,\cos(\phi))$ plane is tangent to the line
$\cos(\phi)=\cos(\alpha)$. This curve goes only through the two upper
regions of the plane, where there
are either no electrons, or the density is pitch angle dependent -
where the integrand in equation \ref{gyroe} is most
likely to be negative.
This special curve also has the longest possible path through the
negative contribution region, and  therefore the absorption 
coefficient derived from integration along this curve should be the largest
negative absorption coefficient possible.
The turnover of the special curve is at energy 
\begin{equation}
\label{gamma-over}
   \gamma=\nu/(s\nu_B)
\end{equation} 
and the frequency which gives this curve is given by the equation 
\begin{equation}
\label{multiD}
{\nu \over {\nu_B}}={s\over\sqrt{1-n_\pm^2\cos^2(\theta)\cos^2(\alpha)}}
\end{equation}
 \par
  The equation \ref{multiD} is our new approximation. 
Our approximation becomes identical with the Melrose and Dulk approximation 
\ref{DM} for the case $n_\pm^2\cos^2(\theta)\cos^2(\alpha) <<1$. It is, however,
more general and is a good approximation for the frequencies of negative absorption,
even when $n_\pm^2\cos^2(\theta)\cos^2(\alpha)$ is large, as the examples
in section \ref{Comparison of Approximations} demonstrate. \par
   Even though we utilize the approximation \ref{multiD} it is important to note that 
it is also possible for a curve of the first form, with $s \nu_B > \nu$,
to give a negative absorption coefficient. This may occur if the curve 
reaches $\cos(\phi)=\cos(\alpha)$ for an energy which is close to the low-energy 
cut-off of the fast electron distribution.
For this case most of the curve at $\cos(\phi) < \cos(\alpha)$ is at energies
where there are no fast electrons. This part of the curve therefore has
no contribution to the absorption coefficient, and since this is the region 
where the positive contribution comes from, the resulting absorption is
negative. 
  We used in our calculations a low-energy cut-off of
$\gamma_{min}=1.02$, which is a common assumption. It turns out that it
is very unlikely that equation
\ref{cosphi_s} gives $\cos(\phi_s)=\cos(\alpha)$ for an energy lower or
close to this cut-off.
  The equation \ref{multiD} therefore gives most of the solutions, and can be 
used to find the frequencies of negative absorption. \par
  Equation \ref{multiD} is still an approximation, and in deriving it we
made simplifying assumptions . For example frequencies close to the special 
frequency given by equation \ref{multiD} also have negative absorption, because
as long as the curve for some frequency has a short path through the
isotropic region the absorption can remain negative.
 Equation \ref{multiD} also assumes a single
harmonic $s$ while the absorption coefficient equation \ref{gyroe}
includes a sum over all harmonics from $s_1$ to $s_2$.
The positive contribution of other harmonics moves the negative
absorption towards zero, and can not be easily approximated. 
Therefore the frequency with largest negative absorption is usually not
exactly the solution of equation \ref{multiD}, but a frequency
close to it. \par
  A close look at equation \ref{multiD} shows that it must be
solved numerically, because the refraction index $n_\pm$ is a function
of the frequency $\nu$. The refraction index is where the ratio
between the plasma frequency $\nu_p$ and the cyclotron frequency 
$\nu_B$ comes into play, since it is very sensitive to this
ratio for low multiples of $\nu_B$. Our approximation is
therefore only semi-analytical, since some numerical computation must
be made. The computation, however, is very simple and can be performed
very quickly. \par
  There are some general properties of the solutions it
is possible to derive even without calculation of the refraction index, and we proceed 
to do so in the next section.  \par
   Finally, equation \ref{multiD} was derived with the assumption that 
$\cos(\theta)>0$, but if $\cos(\theta)<0$ then $\cos(\alpha)<0$ as well, and 
the solution is the same. Equation \ref{multiD} can therefore be
used for all cases. \par

\section{General Properties of the Solution}
\label{general properties}
  The first general property of the solution given in equation
\ref{multiD} is the requirement that $\nu > s\nu_B$.
This may not always be the case (as shown in the previous section),
but it turns out that this condition holds for the dominant (defined as the
most strongly amplified) modes. In practical terms this condition sets the appearance of
negative absorption at frequencies between $s\nu_B$ and $(s+0.5)\nu_B$. \par
  Other general properties can be demonstrated by a graphical solution of
equation \ref{multiD}.  On a plane where the x-axis are the frequency we
plot the line $y=\nu$, and the function
\begin{equation}
\label{curve1}
f(\nu) = { {s\nu_B} \over { \sqrt{1 - n_\pm^2 \cos^2(\theta) \cos^2(\alpha)} } }
\end{equation}
 Any frequency for which the two curves intersect is a solution
of equation \ref{multiD}. \par 
  There are two special cases for the OM and the XO mode
where we can compare $f(\nu)$ and $y=\nu$ without explicitly calculating 
the refraction index. The first case is at the special points of the cut-off 
frequencies of the modes,where $n_\pm=0$. 
The second special case is for large frequencies, where the refraction indices 
are $n_\pm=1$.
The general shape of the curve \ref{curve1} can be inferred
from these two points. \par
  The OM cut-off is at $\nu=\nu_p$, and 
therefore for the Ordinary Mode  $f(\nu_p)=s\nu_B$, 
and if the plasma frequency is equal to the cyclotron frequency, or some
harmonic of the cyclotron frequency, then $\nu_p$ is a solution.
The XO mode cut-off is at the frequency
\begin{equation}
\nu_x=\nu_B/2+\sqrt{\nu_p^2+\nu_B^2/4}
\end{equation}
This cut-off is always at higher frequency than $\nu_B$, and therefore
$f(\nu_x)<\nu_x$ with $s=1$. However, if $\nu_p=\sqrt{2}\nu_B$
then the cut-off is at $\nu_x=2\nu_B$, and the second harmonic $s=2$
is a solution. For any harmonic $s$, the cut-off frequency is equal
to $s\nu_B$ for a ratio $(\nu_p/\nu_B)^2=s^2-s$. \par
\begin{figure}
\resizebox{\hsize}{!}{\includegraphics{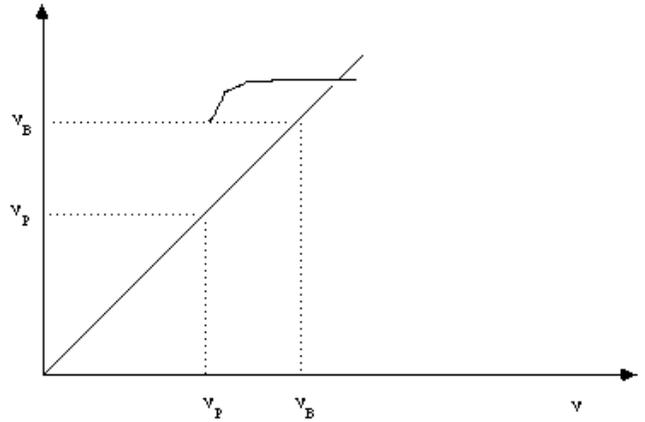}}
\caption[Graphical solution for the frequency]
{A depiction of the plot for $f(\nu)$ (equation \ref{curve1}) for the OM, and the case
$\nu_B > \nu_p$. The curve begins above the line $y=\nu$, and there is
only one possible intersection, and therefore only one possible solution
of equation \ref{multiD}. }
\label{OMabove}
\end{figure}
\begin{figure}
\resizebox{\hsize}{!}{\includegraphics{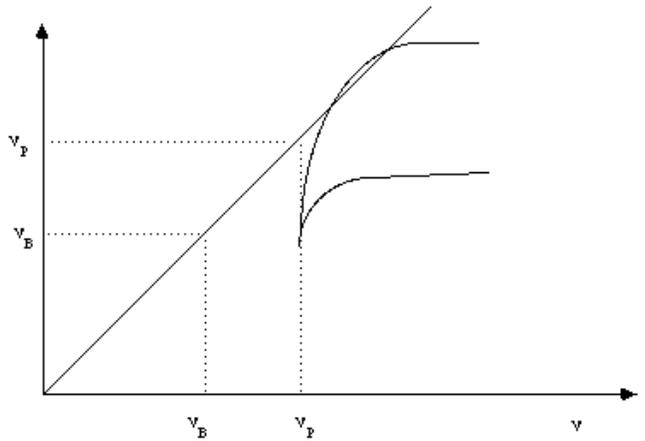}}
\caption[Graphical solution for two frequencies]
{A depiction of the plot for $f(\nu)$ (equation \ref{curve1}) for the OM, and the case
$\nu_B < \nu_p$. The curve begins below the line $y=\nu$, and therefore
there may be no solution, or there are two solutions of equation \ref{multiD}. }
\label{OMbelow}
\end{figure}

  The cut-off frequencies of the modes are not important themselves,
because a wave with a zero refraction index can not propagate.
However, the solutions can be used to deduce the behaviour of the
curve \ref{curve1}. The line $y=\nu$ and the curve
\ref{curve1} can have one of two relations at the cut-off frequencies :
\begin{itemize}
\item If the plasma frequency is smaller than the cyclotron frequency
then $f(\nu_p)=\nu_B > \nu_p$ and the curve \ref{curve1} for the OM
starts above the $y=\nu$ line (figure \ref{OMabove}).
More generally, for all harmonics $s$ such that $s\nu_B > \nu_p$
the curve for the OM starts above the $y=\nu$ line.
\item Conversely, if the plasma frequency is larger than some harmonic $m$
of the cyclotron frequency, then the OM curve \ref{curve1} starts below
the $y=\nu$ line for $s$ up to and including the harmonic $m$
(figure \ref{OMbelow}).
\item If
$(\nu_p/\nu_B)^2 < s^2-s$, then $f(\nu_x)=s\nu_B > \nu_x$, and
the curve \ref{curve1} for the XO starts above the line $y=\nu$.
because then $\nu_x < s\nu_B$.
\item If 
 $(\nu_p/\nu_B)^2 > s^2-s$, then the XO curve starts below 
the line $y=\nu$.
\end{itemize}

  The behaviour of the curve \ref{curve1} for frequencies larger
than the cut-off frequencies is deduced from the refraction index.
Both the OM and the XO mode have refraction indices smaller than $1$
which approach $1$ asymptotically as the frequency increases.
A curve like the curve \ref{curve1} therefore has the shape
illustrated in figure \ref{OMabove} if it  
starts above $y=\nu$. 
From figure \ref{OMabove} it is seen that for a starting point
above the $y=\nu$ line there is only one intersection.
  If the curve starts below the $y=\nu$ line there are either two
intersections, or no intersection at all, as is illustrated in
figure \ref{OMbelow}.
The conclusion is that for the OM there is one solution for negative
absorption coefficients if $s\nu_B>\nu_p$,
and for the XO there is one solution if $\sqrt{s^2-s}\ \nu_B >\nu_p$. 
There are either two solutions or no solution for the OM
if $s\nu_B<\nu_p$, and for the
XO there are two solutions or no solution
if $\sqrt{s^2-s}\ \nu_B<\nu_p$. 
In practice for the cases where the curve \ref{curve1} starts below
the $y=\nu$ line, there is usually no solution, and only the case
illustrated in figure \ref{OMabove} is important. \par
  There is an additional limit we can derive from equation \ref{multiD}.
The mathematical formula derived in equation \ref{curve1} can be plotted
as a function of frequency for any frequency. However, beyond some
frequency the mathematical result $f(\nu)$ is no longer relevant
to finding the negative absorption coefficient.
The most important limit is derived from equation \ref{curve1}
itself, and it is the highest frequency possible for a
given harmonic $s$ with this equation.  Clearly, plotting $f(\nu)$
beyond this frequency is meaningless. The highest possible frequency is
derived for the OM and the XO by assuming that $n_\pm=1$, and
is therefore :
\begin{equation}
\label{nu_max}
  \nu_{max}={{s\nu_B}\over\sqrt{1-\cos^2(\theta)\cos^2(\alpha)}}
\end{equation}
For example with $\cos(\alpha)=0.8$, $s=1$, and emission
direction $\cos(\theta)=0.3$ the upper limit on the frequencies
usable for this set of parameters is $\nu < 1.03 \nu_B$.
If we take $\cos(\theta)=0.7$ the limit is $\nu<1.207 \nu_B$.
This result gives the range of frequencies
where negative absorption is expected. For example, $\cos(\theta)=0.3$
can only have negative absorption between $\nu=\nu_B$ and $\nu=1.03 \nu_B$. \par
 The condition \ref{nu_max} can also be used to limit the
harmonics relevant for any frequency and angle of emission.
For example the first harmonic $s=1$ is not
useful for finding negative absorption coefficients at $\nu > 2 \nu_B$
for any loss-cone with $\cos(\alpha)<0.866$. It is thus possible
to immediately disqualify certain harmonics from contributing
in certain frequency ranges, and shorten the computation
time further. \par
   In summary, we conclude that for the OM and the XO mode, we expect
negative absorption coefficient to appear for frequencies obeying the following
two conditions :
\begin{itemize}
\item For the OM the frequency should be derived from a harmonic $s$ which 
         has $s > \nu_p/ \nu_B$.
\item For the XO mode the frequency should be derived from a harmonic $s$ which 
         has $s^2-s > (\nu_p/\nu_B)^2$.
\item  For both modes the frequency must be smaller than $\nu_{max}$ of equation
          \ref{nu_max}
\end{itemize}
    The harmonic number $s$ of a frequency can be derived from the condition \ref{nu_max}
and should be the smallest integer which fits the inequality 
\begin{equation}
  s  \geq { {\nu}\over{\nu_B}} \sqrt{1-\cos^2(\theta)\cos^2(\alpha)}
\end{equation}
  The largest harmonic $s$ which contributes to a frequency is given by equation
\ref{s1-s2}, but it is reasonable to assume that the smallest harmonic has the
largest contribution, and therefore we can consider it alone for the purposes
of the approximation. \par
   So far we have not considered the important Z-mode (ZM).
The Z-mode is the lower branch of the XO mode, which reappears
at frequencies smaller than $\nu_z$
\begin{equation}
2\nu_z^2=\nu_B^2+\nu_p^2+\sqrt{(\nu_B^2+\nu_p^2)^2-
4 \nu_B^2 \nu_p^2 \cos^2(\theta)}
\end{equation}
and its cut-off frequency is $\nu_x-\nu_B$. \par
Emission in the Z-mode can not emerge from the plasma, and therefore can not be
observed. Since our purpose is to derive estimates for the observable
frequencies, approximate solutions for the Z-mode are not of great interest to us.
However, the Z-mode may be the dominant mode and quench the maser before the 
other modes are amplified. It is therefore of interest to determine where possible
negative frequencies can appear. \par
  The upper Z-mode frequency $\nu_z$ is bound from above by the upper hybrid
frequency
\begin{equation}
\nu_{uh}=\sqrt{\nu_p^2+\nu_B^2}
\end{equation}
  From the equation it is clear that $\nu_z$ decreases with increasing
$\cos(\theta)$, and for $\cos(\theta)=1$ the limiting frequency is
either $\nu_z=\nu_B$ if $\nu_p<\nu_B$ or $\nu_z=\nu_p$ if $\nu_p>\nu_B$.
The lower frequency boundary of the ZM is always smaller than $\nu_p$. \par
  The ZM refraction index is  $n_{ref} > 1$ for frequencies $\nu>\nu_p$, 
is $n_{ref}=1$ at the plasma frequency, and
is smaller than $1$ for frequencies smaller than $\nu_p$. \par
\begin{figure}
\resizebox{\hsize}{!}{\includegraphics{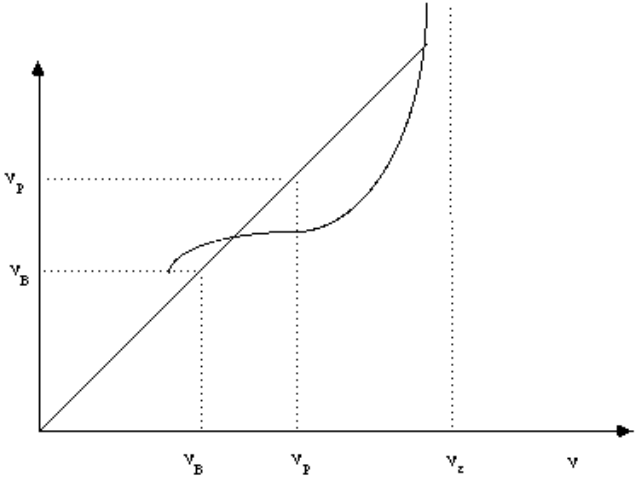}}
\caption[Graphical solution for the Z-Mode]
{A depiction of the plot for $f(\nu)$ (equation \ref{curve1}) for the ZM,
  and the case $\nu_p > s\nu_B/\sqrt{1-\cos^2(\theta)\cos^2(\alpha)}$.
  The curve begins below the line $y=\nu$, and therefore there
  may be one solution for $\nu>\nu_p$ or two solutions, one of
  them for $\nu<\nu_p$.}
\label{ZMgeneral}
\end{figure}

  Using the above characteristics we can determine
the general behaviour of the curve \ref{curve1} for the Z-mode, 
which is illustrated in figure \ref{ZMgeneral}. The curve goes to
infinity for frequencies near $\nu_z$, and goes to $s\nu_B$
as the frequency approaches $\nu_x-\nu_B$. \par
  As for the OM and the XO mode it is possible to deduce the
general behaviour of the solutions for the Z-mode from the
properties of the refraction index, however, because of the additional
dependence on $\cos(\theta)$ the ZM properties are much harder
to pin down.
\begin{itemize}
\item If $\nu_p > \nu_{max}$ (see equation \ref{nu_max}),
  the line $y=\nu$ is above the curve \ref{curve1} at the plasma
  frequency. Since the curve \ref{curve1} goes to infinity near
  $\nu_z$, there is an intersection near $\nu_z$, for a frequency
  $\nu>\nu_p$.
  The effect of $\cos(\theta)$ complicates this
  condition, which is not just $\nu_p > s\nu_B$, since for large
  $\cos(\theta)$ even a high plasma frequency may not be larger
  than $s\nu_B/\sqrt{1-\cos^2(\theta)\cos^2(\alpha)}$.
\item  If in addition to the above also  $\nu_x-\nu_B < s\nu_B$
  then there is also an intersection for $\nu<\nu_p$. 
\item If $\nu_p < \nu_{max}$,
  the line $y=\nu$ is below the line of curve \ref{curve1} at $\nu_p$.
  If, in addition, $\nu_p > s\nu_B$, then there is no solution.
  This condition tells us that there are no ZM solutions for
  large $\cos(\theta)$.
\item  If $\nu_p < \nu_{max}$,
  but $\nu_p < s\nu_B$ than there is no intersection
  in the range $\nu_p < \nu < \nu_{max}$.
  However, if the refraction index changes slowly in the vicinity of
  $s\nu_B$, and $\nu_z$ is relatively far from $s\nu_B$, then it
  is possible that the line $y=\nu$ can intersect the curve \ref{curve1}
  from below. For such a case there is also a second solution, closer
  to $\nu_z$, when the curve \ref{curve1} goes to infinity.
  The condition that $\nu_z >> s\nu_B$ is that $\cos(\theta)$ is small,
  and that $\nu_p$ is not too small.
  For example, if $\nu_p = 0.25 \nu_B$
  then $\nu_z < 1.0308 \nu_B$, and the requirement is that this
  frequency be ''far'' from
  $\nu_B/\sqrt{1-\cos^2(\theta)\cos^2(\alpha)}$. For
  our standard $\cos(\alpha)=0.81$ this means $\cos(\theta)<<0.3$.
  It is reasonable that for even smaller plasma frequencies the
  condition can not be met, and there is no ZM solution.
\item  Finally, if   $\nu_p < s\nu_B$  then there is certainly
  no intersection for $\nu<\nu_p$.
\end{itemize}
  In summary the condition for negative absorption in the Z-mode is
either $\nu_p > s\nu_B$, or $\cos(\theta)<<1$ and $\nu_p/\nu_B$
not too small. Empirically ''not too small'' translates as
$\nu_p/\nu_B \geq 0.25$. \par
  An interesting phenomena for the Z-mode negative absorption is that for
$1.4 < \nu_p/\nu_B < 1.8$ the frequency range of the Z-mode is between
$\nu_B$ and $2\nu_B$. However, for this range of 
plasma frequencies the energy of the turnover \ref{gamma-over} is relatively
high for $s=1$, and it turns out that most of the curve described
by \ref{cosphi_s} passes through regions of few electrons. The result is
that the negative absorption coefficient are very small in absolute value,
or that there are no negative absorption coefficients at all for
these $\nu_p/\nu_B$ ratios.  \par

\section{Comparison of Approximations}
\label{Comparison of Approximations} 
\begin{figure}
\resizebox{\hsize}{!}{\includegraphics{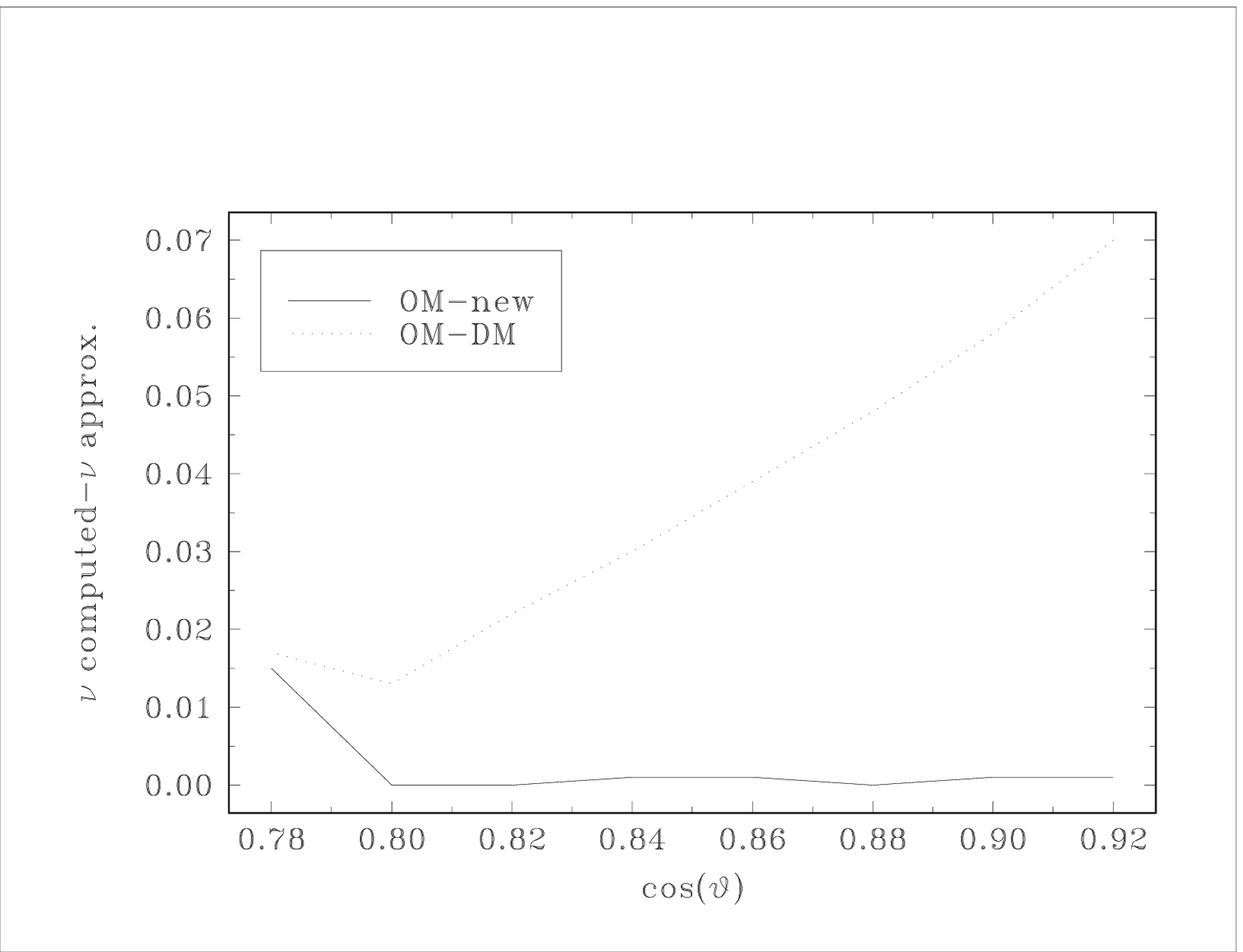}}
\caption[Standard parameters first harmonic]
{The difference between the frequency of largest in absolute magnitude
negative absorption coefficient found by numerical calculations and the frequency
found by the Dulk-Melrose approximation \ref{DM}, and the difference between
the calculation and the new approximation \ref{multiD}, as function of
$\cos(\theta)$, the cosine of emission angle to the magnetic field. 
The calculation is for the standard parameters power law index $\delta=3$,
loss cone angles $\cos(\alpha)=0.81, \cos(\alpha-\delta\alpha)=0.83$, magnetic
field of $B=360 \ gauss$, and plasma frequency $\nu_p=\nu_B$.
 The frequencies are for the first harmonic between $\nu_B$ and $2\nu_B$.}
\label{100-81-1}
\end{figure}

\begin{figure}
\resizebox{\hsize}{!}{\includegraphics{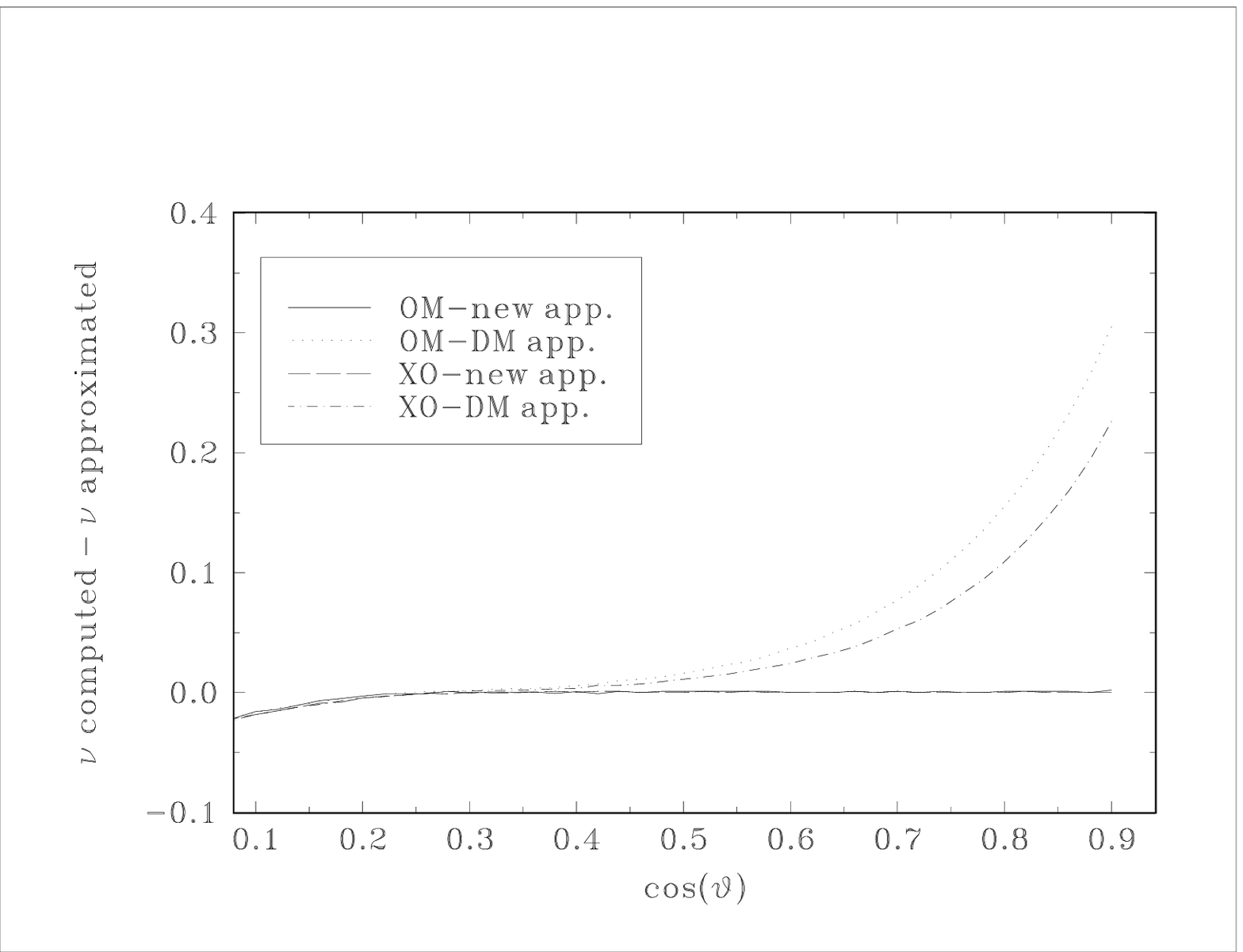}}
\caption[Standard parameters second harmonic]
{The difference between the frequency of largest in absolute magnitude
negative absorption coefficient found by numerical calculations and the frequency
found by the Dulk-Melrose approximation \ref{DM}, and the difference between
the calculation and the new approximation \ref{multiD}, as function of
$\cos(\theta)$, the cosine of emission angle to the magnetic field. 
The calculation is for the standard parameters power law index $\delta=3$,
loss cone angles $\cos(\alpha)=0.81, \cos(\alpha-\delta\alpha)=0.83$, magnetic
field of $B=360 \ gauss$, and plasma frequency $\nu_p=\nu_B$.
 The frequencies are for the second harmonic between $2\nu_B$ and $3\nu_B$.}
\label{100-81-2}
\end{figure}

\begin{figure}
\resizebox{\hsize}{!}{\includegraphics{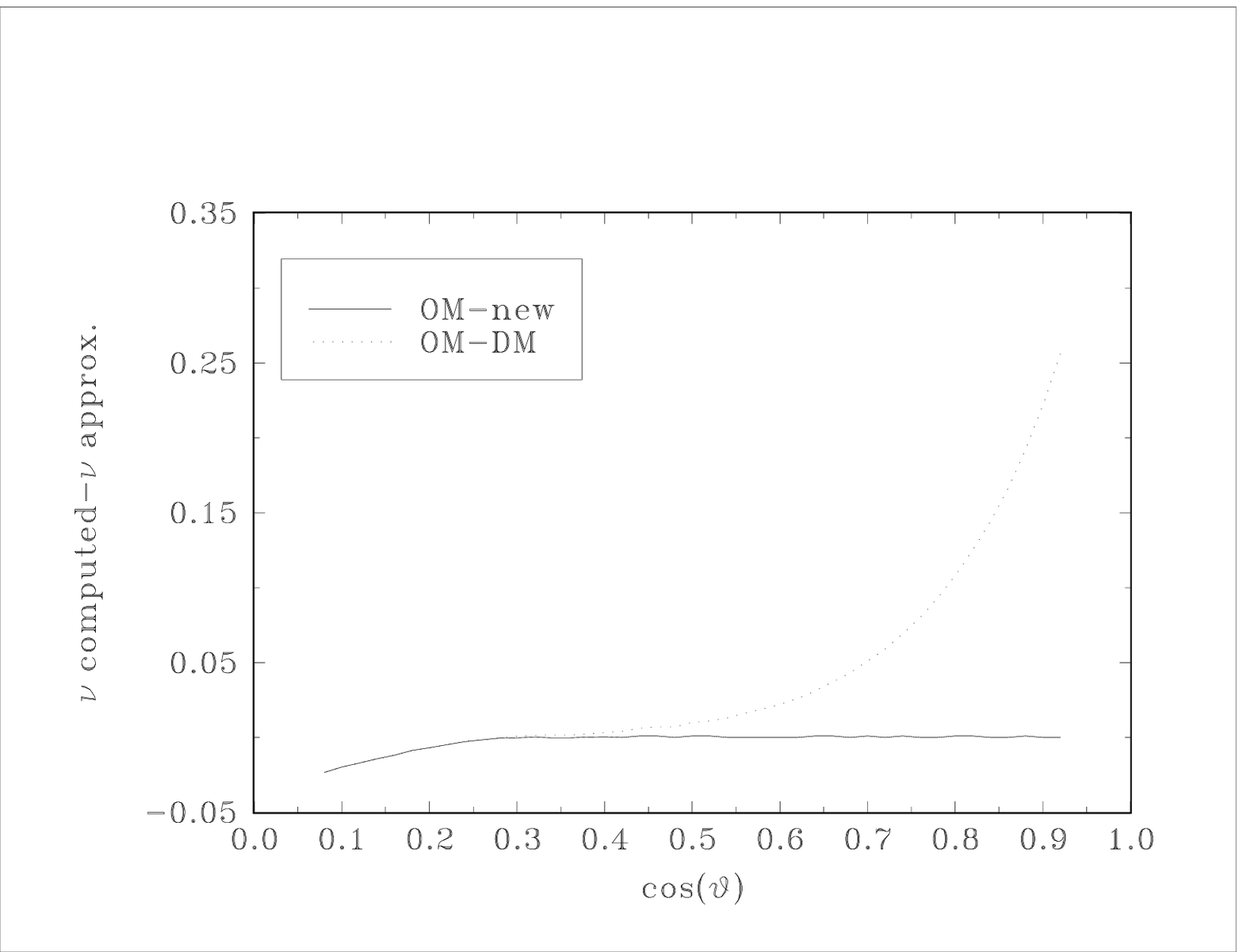}}
\caption[Large plasma frequency]
{The difference between the frequency of largest in absolute magnitude
negative absorption coefficient found by numerical calculations and the frequency
found by the Dulk-Melrose approximation \ref{DM}, and the difference between
the calculation and the new approximation \ref{multiD}, as function of
$\cos(\theta)$, the cosine of emission angle to the magnetic field. 
The calculation is for the standard parameters power law index $\delta=3$,
loss cone angles $\cos(\alpha)=0.81, \cos(\alpha-\delta\alpha)=0.83$, magnetic
field of $B=360 \ gauss$, and for a large plasma frequency $\nu_p=1.4 \nu_B$.
 The frequencies are for the second harmonic between $2\nu_B$ and $3\nu_B$.}
\label{140-81-2}
\end{figure}

  We performed numerical calculations of the absorption coefficient
\ref{gyroe} using the power law distribution \ref{powerlaw}
with index $\delta=3$, and the loss 
cone distribution \ref{losscone} with $\cos(\alpha)=0.81,
\cos(\alpha-\delta\alpha)=0.83$.
The calculation were performed with a standard magnetic field
$B=360\ gauss$, and an ambient number density which was changed
to give different ratios of $\nu_p/\nu_B$.
For every cosine of emission angle $\cos(\theta)$ to the magnetic field
the largest in absolute
magnitude negative absorption coefficient was found
by scanning in frequency from $s\nu_B$ to $(s+1)\nu_B$.
The frequency of this largest in absolute magnitude negative absorption
coefficient is compared with the approximation \ref{DM} and with
our new approximation \ref{multiD} in the figures \ref{100-81-1}
to \ref{140-81-2}. \par
  In figure \ref{100-81-1} the comparison is made for frequencies
between $\nu_B$ and $2\nu_B$, and the figure shows that for small
$\cos(\theta)$ the Dulk-Melrose approximation \ref{DM} is very similar
to our new
approximation. However, for larger $\cos(\theta)$, the new approximation
is much better. The new approximation is identical with the
result of the full numerical computation for most of the $\cos(\theta)$
range where there is negative absorption. \par
  In figure \ref{100-81-2} the comparison is made for frequencies
between $2\nu_B$ and $3\nu_B$. Here the relevant $\cos(\theta)$ range
is larger, and again for small cosines the Dulk-Melrose approximation
\ref{DM} is similar to the numerical results and to our new
approximation. However, for angles smaller than about $60$ degrees,
the Dulk-Melrose approximation begins to diverge, while the new
approximation remains virtually identical with the numerical
results. \par
  For the ratio $\nu_p=1.4\nu_B$ our conclusions in section
\ref{general properties} lead us to expect that negative absorption
exists only for frequencies $\nu > 2\nu_B$, and the numerical
computations bear this out.
  Figure \ref{140-81-2} shows again that for large $\cos(\theta)$
the approximation of equation \ref{DM} begins to diverge from
the numerical results, while our new approximation remains
identical to it. \par
  We do not show results for the Z-mode, since we are interested in presenting
estimates for the frequencies of occurrence of observable emission.
The Z-mode is important in quenching the maser, but it is necessary to
compute the absorption coefficients for all the modes and compare them to
determine whether it does. \par

\section{Discussion}
  We develop a new approximation for the frequencies at which
the absorption coefficient is negative, and therefore the Electron
Cyclotron Maser mechanism operates. This new approximation is
given by equation \ref{multiD}, and is easy to compute.
Our approximation gives results which are much more accurate than 
previous approximations, and are practically the same as the results
of a full numerical calculation. The frequencies derived with the approximation
are within $0.01-0.02\ \nu_B$ of the numerically calculated frequencies. \par
  The parameters entering the approximation are the angle of emission to the
magnetic field $\theta$, the loss-cone opening angle $\alpha$,
the ratio $\nu_p/\nu_B$, and the ratio $\nu/\nu_B$. \par
  The new approximation can be used to define a range of possible
frequencies of millisecond spike emission, given the ratio
$\nu_p/\nu_B$. Or, when spike emission is detected, the approximation
can be used to give the range of physical parameters in the emission
region.

\clearpage

\end{document}